\documentclass{aa}
\usepackage{graphics,epsf,epsfig,graphicx}
\begin{document}

%
   \title{Mapping the submillimeter spiral wave in NGC 6946}

   \subtitle{}

   \author{P.B.Alton\inst{1}
	\and S.Bianchi\inst{2}
	\and J.Richer\inst{3}
	\and D.Pierce-Price\inst{3}
	\and F.Combes\inst{4}
}

   \offprints{P.B.Alton, paul.alton@astro.cf.ac.uk}

   \institute {Department of Physics \& Astronomy, University of Wales, PO Box 913, 
Cardiff CF2 3YB, U.K.
   \and ESO, Karl-Schwarzschild-Strasse 2, D-85748 Garching bei Muenchen, Germany.
   \and Cavendish Radio Laboratory, Madingley Road, Cambridge.
   \and LERMA, Observatoire de Paris, 61, Avenue de l'Observatoire, 75014 Paris, France.
}

   \date{Received 2001; accepted 2002}
\authorrunning{Alton et al.} 
\titlerunning{Submillimeter map of NGC 6946}

   \abstract{
We have analysed SCUBA $850\mu$m images of the (near) face-on spiral galaxy NGC 6946, 
paying particular attention to the subtraction of sky signal. A comparison with both 21cm HI 
and $^{12}$CO(2-1) intensity maps reveals a tight correlation between dust thermal emission 
and {\em molecular} gas at the kiloparsec level. By means of a Monte Carlo radiative transfer 
model, we convert a B-K colour image of NGC 6946 into a map of visual optical depth. The 
model yields maximum opacities since we assume that any increase in B-K colour, with 
respect to the disk edge, is attributable solely to extinction by dust. The resultant map of 
visual optical depth relates well to the distribution of neutral gas (HI+H$_{2}$) and implies 
a global gas-to-dust ratio of 90 (this value 
is a lower limit). There is no significant radial variation of
this ratio: this can be understood, since the gas content is dominated by far by the molecular
gas. The latter is estimated through the CO emission tracer, which is itself dependent on
metallicity, similarly to dust emission. In the absence of a more objective tracer,
it is not possible to derive the true gas-to-dust ratio.
By comparing the radial profile of our visual optical depth map 
with that of the SCUBA image, we infer an emissivity (dust absorption coefficient) at 
$850\mu$m that is 3 times lower than the value measured by COBE in the Milky Way,
and 9 times lower than in NGC 891. 
We view this very much as a lower estimate, however, given our initial assumptions in deriving 
the visual opacity, 
 and the possibility of underestimating the large-scale submm emission,
the effect being more severe for the nearly face-on orientation of NGC 6946.
A decomposition of the spiral structure half way out along the disk of 
NGC 6946 suggests an interarm optical depth of between 1 and 2. These surprisingly high 
values represent 40-80\% of the visual opacity that we measure for the arm region.
}
   \maketitle


%

\section{Introduction}
\label{intro}

Modern submillimeter (submm) and far-infrared (FIR) facilities such as the Submillimeter Common 
User Bolometer Array on JCMT (SCUBA), 
Institut de Radioastronomie millim\'{e}trique (IRAM) and the Infrared 
Space Observatory (ISO) have undertaken an invaluable r\^{o}le in re-evaluating the radial 
distributions and global masses of galactic dust in nearby spiral galaxies (e.g.~Gu\'{e}lin et al 
1993; Gu\'{e}lin et al 1995; Neininger et al 1996; Alton et al 1998a; Siebenmorgen et al 1999). 
Nevertheless, detailed information about how grains are distributed with respect to the spiral 
arm structure, or how they relate at the kiloparsec level to the various gas phases (HI, H$_{2}$) 
has been less forthcoming (Knapen \& Beckman 1996; Seigar \& James 1998; Rand et al; 1999; Alton 
et al 2001). A major obstacle, in this respect, is the limited spatial resolution available with 
instruments such as ISO ($\sim 100\arcsec\,$ at $200\mu$m) -- barely sufficient to 
resolve the spiral 
arms in even the closest galaxies. At wavelengths closer to 1 mm, ground-based arrays such as 
SCUBA and IRAM, afford considerably better spatial resolution ($\sim 10\arcsec\,$) but 
emission arises 
from the Rayleigh-Jeans tail in this regime and, consequently, surface brightnesses are very low 
(usually $\sim 10$ mJy/beam). The sparse emission is particularly problematic for face-on spirals 
and it is precisely galaxies with this kind of inclination that we need to study if we are to 
understand how gas and dust relate to the spiral density wave.\\

At the {\em global} level several important trends have emerged from recent submm/mm observations 
of spirals. For example, Dunne et al (2000) have shown that the total $850\mu$m flux density 
correlates more tightly with integrated CO(1-0) line emission than with the 21cm HI line emission.
The mid-IR emission (7 to 15 microns) detected with ISOCAM does not correlate
well with HI and CO either (e.g. Genzel \& Cesarsky, 2000,  Frick et al 2001).   
Total dust masses that include submm/mm data now begin to show signs that they are comparable 
to dust masses inferred from optical extinction/reddening (Block et al 1994; Alton et al 1998b). 
Nevertheless, at the {\em local} level ($\sim1$ kpc) it is less clear how grain thermal emission 
relates to optical absorption lanes, star-forming regions, the diffuse interstellar medium (ISM) 
and, of course, the various gas phases (HI,H$_{2}$,HII).\\

In a recent paper, Alton et al (2001) collated and examined the most up-to-date estimates of 
spiral disk opacity. What became obvious in this study is that relatively little is known about 
how dust is distributed with respect to the spiral arms and, in particular, whether interarm 
grain material is prevalent or not. The interarm region is particularly important because it 
occupies more than 50\% of the disk and may therefore be crucial in determining 
the ability of spiral disks to block background light (Heisler \& Ostriker 1988; Fall \& Pei 1993; 
Gonzalez et al 1998; Masci \& Webster 1999). Furthermore, the distribution of dust with respect 
to the spiral density wave may hold clues as to how gas is compressed along the spiral arms to 
form new stars. We should not forget that half of the metals in spiral galaxies are 
encapsulated in grains and consequently galactic dust provides a valuable `fossil record' of 
previous star-formation within the disk environment (Whittet 1992; Williams \& Taylor 1996; 
Edmunds \& Eales 1998).\\

\begin{table}[h]
\centering
\caption{Basic properties of NGC 6946.}
\label{tab1}
\begin{tabular}{lll}

Parameter 	& Value				& Reference \\ \\
Type 		& Sc 				& de Vaucouleurs et al (1991) \\
R$_{25}$ radius	& $5.45\arcmin\,$			&  	\hspace{0.6in} " \\
R.A.(J2000) 	& $20^{h} 34^{m} 52^{s}.3$ 	&  	\hspace{0.6in} " \\
Dec. (J2000) 	& $60^{\circ} 09\arcmin\, 14.2\arcsec\,$ 	& 	\hspace{0.6in} " \\
Distance 	& 10 Mpc 			& Tacconi \& Young (1986) \\
Inclination 	& $30^{\circ}$ 			& Corradi \& Capaccioli (1991) \\
P.A. 		& $63^{\circ}$ 			& 	\hspace{0.6in} "	\\
\\
\end{tabular}
\end{table}

In this paper, we address some of the issues outlined above by analysing data for NGC 6946. 
In most respects, this nearby (10 Mpc) galaxy resembles a typical Sc spiral and is oriented 
favourably for a study of face-on spiral structure (table~\ref{tab1}). The centre of NGC 6946 
evidently contains a large amount of recent star-formation (Tacconi \& Young 1990; Engargiola 1991). 
However, there is no evidence from optical and near-infrared (NIR) imaging of tidal disturbance 
and indeed, the disk appears quite isolated on the sky
(except for the low surface brightness spiral Cep 1, at 0.4 Mpc in projected
distance, Burton et al. 1999). NGC 6946 is peculiar for its richness in CO emission,
the derived molecular-to-atomic gas mass being quite high (Casoli et al, 1990).
The density waves are not very regular nor contrasted, and in particular
the total radio continuum emission  is only slightly enhanced in spiral arms,
indicating low compression (Boulanger \& Viallefond, 1992).  The
contrast is even lower, when short spacings are added to interferometric
maps (Frick et al 2001). 
Also, polarized radio emission is concentrated
in interarm regions (Beck \& Hoernes 1996). \\
The global FIR-to-blue luminosity ratio 
is close to unity which is characteristic of quiescent spirals rather than an object undergoing 
extensive starburst activity (Soifer et al 1987; Bianchi et al 2000a). Our submm data for NGC 6946 
embrace a re-analysis of $850\mu$m data collected by Bianchi et al (2000b) as well as some 
additional mapping observations, at the same wavelength, accessed from the SCUBA archive. We 
begin by examining the relationship between submm emission and neutral gas (HI and H$_{2}$; 
\S\ref{submm_gas}). A Monte Carlo radiative transfer model is then applied to a B-K colour map 
of NGC 6946 with a view to comparing visual optical depth with dust thermal emission 
(\S\ref{reddening}). The latter process leads to an upper limit in the grain emissivity at 
$850\mu$m (\S\ref{inferred_emissivity}). Visual optical depth is then compared to column 
density of neutral gas in order to discover how the gas-to-dust ratio changes with galactocentric 
radius and gas-phase metallicity (\S\ref{gd}). Finally, we examine the distributions of HI, H$_{2}$, 
submm thermal emission and visual optical depth with respect to the spiral density wave, providing 
estimates of opacity and gas column density for both arm and interarm regions in NGC 6946 
(\S\ref{arm_interarm}).

\section{Observational Details}
\label{obs}

Our submm data for NGC 6946 stem from two sources: 22 hours of exposure taken by ourselves during 
April and June 1998, and a further 5 hours of observations retrieved from the SCUBA archive. The 
former have been presented briefly in Bianchi et al (2000b). All the data were reduced in a similar 
fashion to that described in Bianchi et al but more attention was now given to the subtraction of the
 sky signal (which is strong with respect to the source signal). The Submillimeter Common User 
Bolometer Array (SCUBA) is mounted at the Nasmyth focus of the James Clerk Maxwell Telescope (JCMT)
 and provides simultaneous imaging at 450 and $850\mu$m by virtue of a dichroic beamsplitter (Holland
 et al 1999), with a field of view of 2.3\arcmin\,.
 The shortwave array ($450\mu$m) consists of 91 bolometers (HPBW=$7.5\arcsec\,$) whilst the 
longwave array ($850\mu$m) is composed of 37 elements (HPBW=$14.7\arcsec\,$). At the shorter wavelength, 
we detect little more than the NGC 6946 nucleus and, therefore, our results will concentrate chiefly
 on the $850\mu$m emission. \\

In scan-map mode, the telescope scans the source at a rate of 24 arcsec per second and at an angle 
allowing the source to be fully sampled (Pierce-Price et al 2000). The scan length was 9\arcmin\,, and
spacing $\sim$ 1\arcmin\,. During the scan, the secondary 
chops at a frequency of 7.8 Hz within the observed field facilitating the removal of sky background
 emission. Each scan-map is a convolution of the source with the chop and the profile of the source
 is eventually restored by a fourier transform (FT) technique. Scan-maps presented here cover most 
of the NGC 6946 disk ($D_{25}\simeq 10\arcmin\,$) and an individual scan took typically 6 minutes. In order
 to sample emission over various scale-sizes, we adopted chop throws of 20, 30 and $65\arcsec\,$ along both
 the RA and Dec directions. \\

The dedicated SCUBA software package, SURF (Jenness 1997), was employed to clean, flatfield and 
calibrate the images as well as correct for atmospheric attenuation. Since the chop takes place within
 the source field, the software also uses the edges of each scan to remove a linear baseline 
corresponding to the sky. At $850\mu$m, the zenith optical depth was usually between 0.1 and 0.3. 
Scan-maps of Venus, processed in the same manner as the galaxy data, provided a calibration from Volts 
to Janskys for each night's observations. The uncertainty in this calibration is estimated to be 15\%.\\

The Emerson II technique enabled a final image to be produced from the chop-convolved data 
(Emerson 1995; Jenness et al 1998; Holland et al 1999). Scan-maps associated with each chop 
configuration were combined into a single image. Each of these 6 maps was then divided by the
 FT of the chop (a sine wave) in order to obtain the FT of the source. The FTs of the source were 
combined before the inverse FT process was carried out to produce the final image. Unfortunately,
 frequencies close to zero, corresponding to slow changes in the sky signal, are poorly sampled by
 this technique. This inadequacy is manifested by large-scale undulations in the convolved images 
for each chop configuration (Pierce-Price et al 2000). The difficulty was alleviated to some extent
 by adopting an iterative approach in the data processing. The final image at $850\mu$m was fed back
 into an earlier stage of the reduction in order to help separate sky fluctuations from object signal.
 In our case, this was repeated 4 or 5 times before no further change was noted in the final image.
 Even after this process, some fluctuations of scalesize $\sim 5\arcmin\,$ were still apparent in the final 
image. Attempts to remove these fluctuations by fitting a polynomial surface or using an unsharp-mask
 technique were largely successful and gave very similar results. We estimate 
that the largest detectable angular scale is of order 5\arcmin\,. A diffuse emission with a flat
radial distribution larger than 5\arcmin\,, the radius of NGC 6946, would not be detected.
In the end, an unsharp-mask technique
 was selected. Here, the original image was smoothed with a gaussian of size $5.5\arcmin\,$ FWHM and this 
smoothed version subtracted from the original version in order to give the signal above the "local"
 zero point. The resultant image was adopted as our final image at $850\mu$m. The $1\sigma$ random 
noise in this final image is estimated to be 4.2 mJy/$16\arcsec\,$beam. The additional 
uncertainty associated
 with the sky subtraction (in particular the frequencies close to zero) is believed to be between 1.0
 and 1.5 mJy/$16\arcsec\,$ beam. This systematic error was estimated by dividing the scan-maps 
up into 3 sets
 (each set taken on different dates), processing each data set separately and comparing the 3 resultant
 final images photometrically. Similarly, various trials in the sky-subtraction process (polynomial
 surface, unsharp-masking etc.) suggest uncertainties of order 1 mJy/$16\arcsec\,$ beam in establishing the 
zero baseline. \\

For comparison with the SCUBA data, a $^{12}$CO(2-1) image of NGC 6946 was kindly made available to 
us by Sauty et al (1998). These data were taken with the IRAM 30-m telescope and have a spatial 
resolution of $13\arcsec\,$. 
 A new CO(1-0) map made with the 30-m telescope with 22\arcsec\, resolution
will be soon available (Walsh et al, in press) and 
is discussed in Frick et al (2001). It is however very similar to that
from Sauty et al (1998).
Nieten et al. (1999) detected CO(4-3),
which indicates the presence of warm gas, as far as the outer spiral arm.
Bianchi et al (2000b) have already noted a strong correlation between the 
$850\mu$m dust thermal emission and the $^{12}$CO(2-1) emission line in NGC 6946. However, the 
relationship is not completely linear and there is also evidence for a bifurcation in scatter-plots 
of two quantities (discussed below). A very extensive image of 
NGC 6946 ($26\arcmin\, \times 26\arcmin\,$), taken 
in the 21cm atomic hydrogen line, has been made available to us more recently by F.Boulanger. These 
data have a spatial resolution of $25\arcsec\,$ and combine both single-dish and interferometer data 
(Boulanger \& Viallefond 1992). Deep images in the B and K wavebands, taken with the KPNO 0.9-m 
and 2.3-m WIRO telescopes, were obtained in April 1996 and November 1995 respectively 
(Trewhella 1998). The K-band image was obtained by means of the $128 \times 128$ pixel Michigan 
Infrared Camera and covers an area of $7.4\arcmin\, \times 7.4\arcmin\,$.

\begin{figure}[t]
\centering
\resizebox{\hsize}{!}{\includegraphics{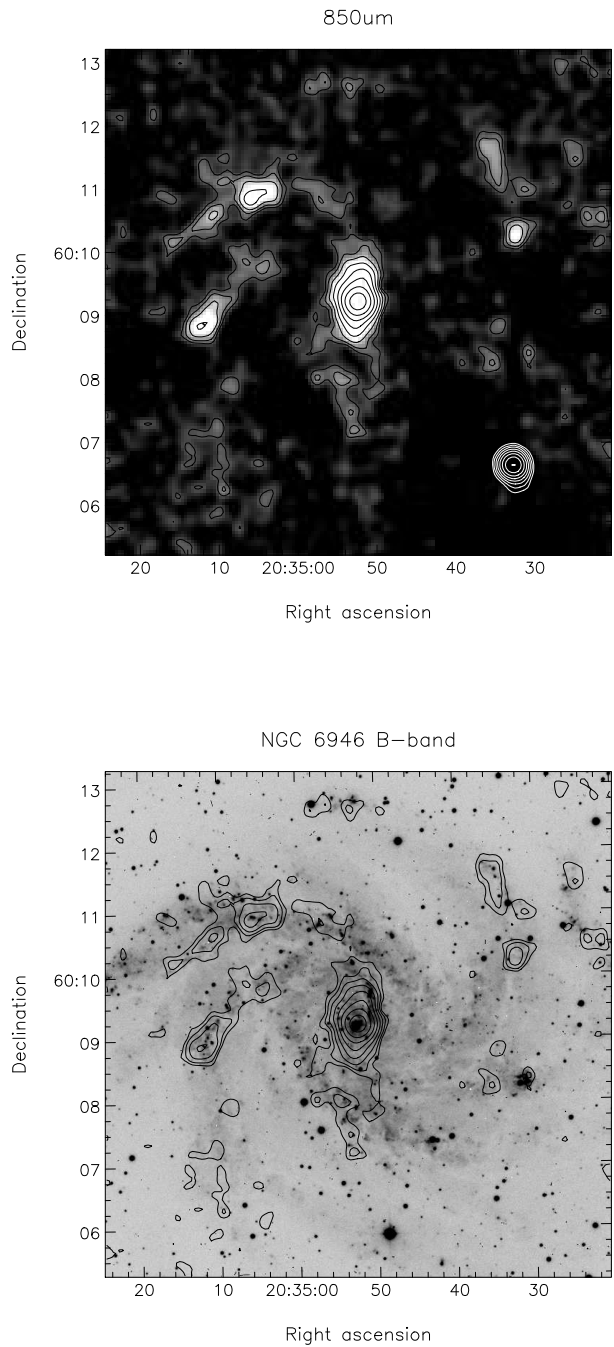}}
\caption{SCUBA $850\mu$m greyscale image (top) superimposed by $850\mu$m contours. Isophotes start
 at $3\sigma$ and are separated by 0.5 mag. The beam ($18\arcsec\,$ FWHM) is shown with white contours at 
the bottom-right of the plotted region. In the bottom figure, the submm isophotes have been projected
 onto a B-band image.}
\label{submm_optical_all}
\end{figure}

\section{Relationship between submm, CO and HI emissions}
\label{submm_gas}

The submm emission is a function of dust column density and dust temperature,
while the HI and CO emissions are function of gas column density and gas excitation.
All three can be considered as gas tracers, with assumptions on gas-to-dust ratios, dust temperature
and gas temperature and density.

The $850\mu$m emission detected from NGC 6946 constitutes almost entirely thermal emission 
from dust grains (at T $\sim$ 15K)
heated by stellar radiation (Israel et al 1999). This assumption begins to 
break down for the central $60\arcsec\,$ of the disk where contamination by the $^{12}$CO(3-2) line 
($\lambda=870\mu$m) to the SCUBA longwave filter might be as high as 33\% (Mauersberger et al
 1999; Dumke et al. 2001). In fig.~\ref{submm_optical_all}, we show the SCUBA $850\mu$m 
image after smoothing to a spatial resolution of $18\arcsec\,$ (FWHM). The salient features of this 
image are similar to those described in detail by Bianchi et al (2000b). The nuclear region is
 a strong source of submm radiation, with a maximum surface brightness of 250$\pm$40 mJy/$18\arcsec\,$beam 
at $850\mu$m. There are also local `hot-spots' situated at $\simeq 2.5\arcmin\,$ away from the nucleus and
 position angles of $44^{\circ}$, $99^{\circ}$ and $203^{\circ}$. These coincide with prominent HII
 regions in B-band and H$\alpha$ images of NGC 6946. The inner 
disk ($2.5\arcmin\, \times 2.5\arcmin\,$) is reasonably
 well detected in our SCUBA image.\\

To compare, on equal terms, the distributions of neutral gas and dust thermal emission, we 
smoothed the $^{12}$CO(2-1) and $850\mu$m images to the same spatial resolution as the 21cm HI map
 ($25\arcsec\,$) using a circular gaussian filter. Photometric sampling was then carried out, on all 3 
images, for those regions where significant $850\mu$m emission ($\ge 2\sigma$) had been detected.
 This was conducted using a $25\arcsec\,$ diameter aperture for a total of 46 regions. To convert the 
$^{12}$CO(2-1) line emission to molecular hydrogen column density, N(H$_{2}$), we adopted 0.4 as
 the ratio of CO(2-1) to CO(1-0) intensity (as observed by Casoli et al 1990 in the main disk). 
We then assumed $X=1.5 \times 10^{20}$cm$^{-2}$ K kms$^{-1}$ for the conversion of CO(1-0) to 
N(H$_{2})$. We recognize sizeable uncertainties in our conversion parameters, in particular a 
factor 2 in the adopted `X' factor. More insidious perhaps is the expected {\em radial} change
 in `X' which is probably as large as any uncertainty in the global value (Neininger et al 1996;
 Dumke et al 1997). Furthermore, the ratio of CO(2-1) to CO(1-0) intensity, within the $25\arcsec\,$ of
 NGC 6946, has been shown to be closer to unity rather than the value of 0.4 adopted here 
(Weliachew et al 1988; Casoli et al 1990). Fortunately, many of the inference we make in this paper
 are based on the main disk and {\em exclude} the central $1\arcmin\,$ of the galaxy. It is this region 
which is likely to show dramatic changes in the conversion factors. Finally, we emphasize that we
 have assumed a fairly low value for $X$, consistent with recent gamma ray studies and the latest
 review articles of the CO(1-0) to N(H$_{2})$ conversion factor (Maloney 1990).\\

\begin{figure}[h]
\centering
\resizebox{6.3cm}{!}{\includegraphics{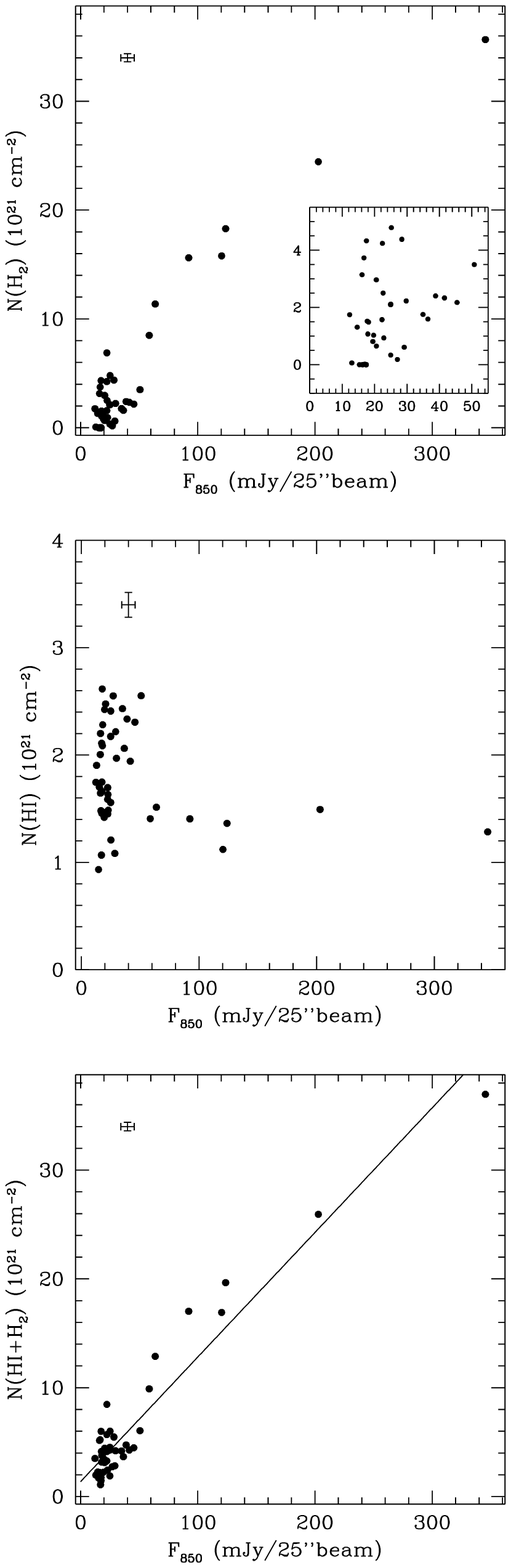}}
\caption{Relationship between neutral gas and $850\mu$m emission in NGC 6946. The submm surface 
brightness is denoted by $F_{850}$ whilst $N(HI)$ and $N(H_{2})$ refer to the column density of 
atomic and molecular hydrogen respectively. In all cases, an error-bar at the top-left of the 
plotted region indicates the random error. The inset in the top graph magnifies data points 
occurring at lower $850\mu$m surface brightness.}
\label{gas}
\end{figure}

Fig.~\ref{gas} demonstrates how the submm emission relates to the hydrogen column density 
derived from CO and 21cm HI maps. The top graph indicates a strong positive trend between molecular
 gas and $850\mu$m surface brightness, $F_{850}$. 
Fig.~\ref{gas} also suggests little correspondance between atomic gas and the submm emission we 
detect. For the disk as a whole this is intuitively correct because the $850\mu$m flux density
 is clearly concentrated towards the centre whereas NGC 6946, like so many other spirals, is known
 to possess a central depression in HI column density (Boulanger \& Viallefond 1992). Including the 
atomic gas in the total hydrogen column density does little to strengthen the correlation between
 neutral gas and $F_{850}$, primarily due to the dominance of N($H_{2}$) at higher submm surface
 brightnesses. A least squares fit between total column of neutral hydrogen, N(HI+H$_{2}$), and 
$F_{850}$ yields the following:

\begin{equation}
\label{eq1}
\frac{N(HI+H_{2})}{10^{21}} = (0.115\pm 0.0052) \times F_{850} + (1.3\pm 0.4)
\end{equation}

\noindent where N(HI+H$_{2}$) and $F_{850}$ are expressed in atoms/cm$^{2}$ and mJy/$25\arcsec\,$beam, 
respectively. 
The two fitted coefficients for the correlation with  N($H_{2}$) alone are
$0.117\pm 0.0057$ for the slope and $-0.5\pm 0.4$ for the zero point.
 The addition of HI slightly improves the correlation, but the
effect is not very large, given the low HI column densities, with
respect to the derived H$_2$ ones.
In order to derive a gas-to-dust ratio from this relationship we are first compelled
 to relate $F_{850}$ to the optical depth of dust in NGC 6946. This step is carried out in the next 
section.

\section{Relationship between submm and reddening}
\label{reddening}

\begin{figure}[t]
\centering
\resizebox{\hsize}{!}{\includegraphics{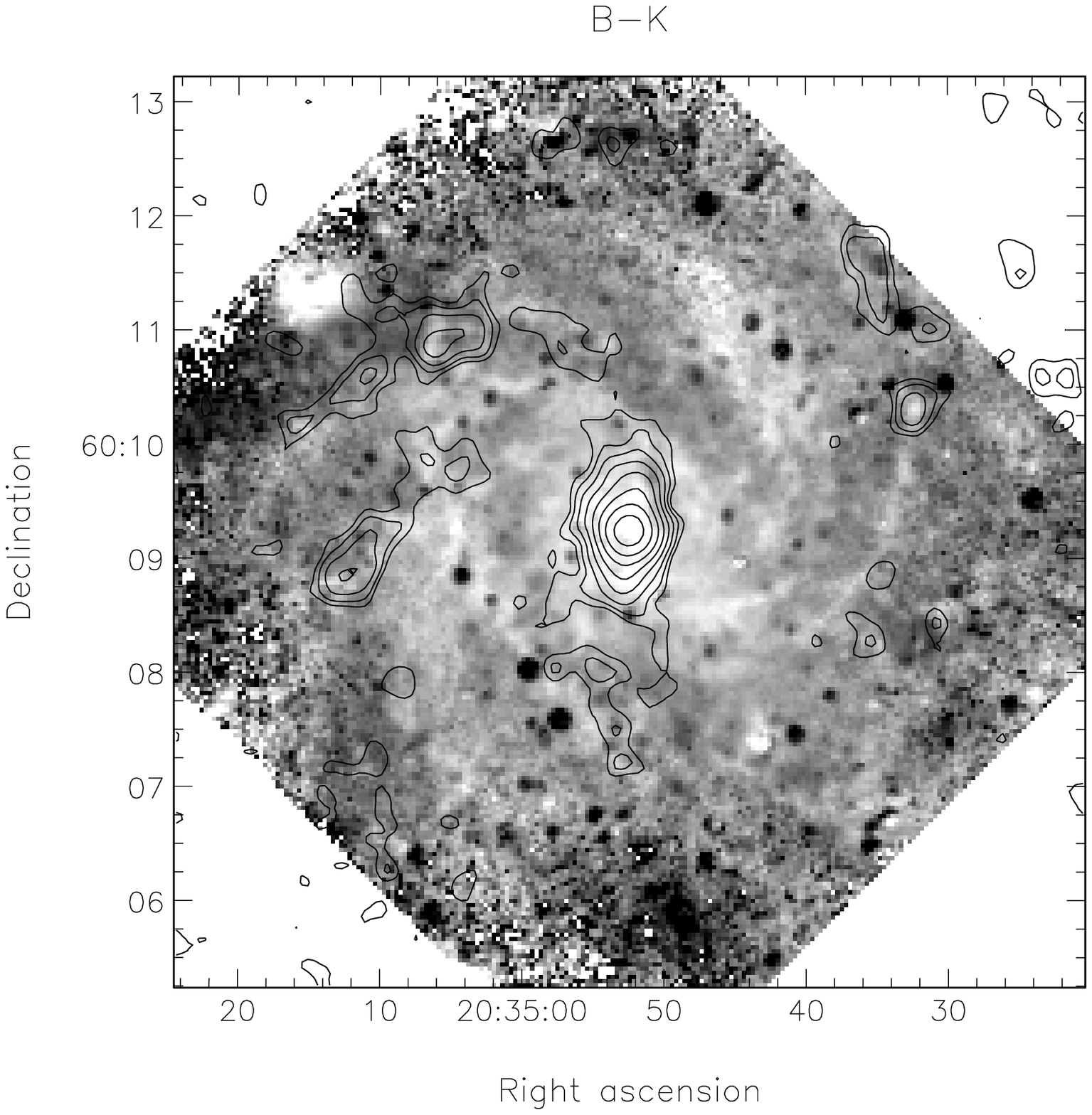}}
\caption{B-K colour image of NGC 6946 displayed in greyscale (reddened regions appear as white).
This is quite similar to the extinction map of Trewhella (1998) combining optical
and far-infrared.
The $850\mu$m contours appearing in fig.~\ref{submm_optical_all} have been superimposed.}
\label{colour_submm}
\end{figure}

In this section, we relate thermal emission from dust at $850\mu$m to reddening evident 
in optical images of NGC 6946. The deep B- and K-band images, alluded to in \S\ref{obs}, were 
aligned and convolved to a common spatial resolution of $4.2\arcsec\,$ and the two images divided to 
produce a B-K colour image in magnitudes. In fig.~\ref{colour_submm} we show this reddening 
image with $850\mu$m contours superimposed. NGC 6946 is typical of a `grand design' spiral 
galaxy with rather prominent spiral dust lanes (Trewhella 1998). We note the presence 
of extensive reddening $\sim 2\arcmin\,$ north-west of the nucleus, while
our 21cm HI and CO images give the impression of little 
atomic and molecular gas at this location. The filamentary structure in this part of the 
B-K map is more suggestive of reddening by dust rather than an absence of blue stars. \\

Fig.~\ref{colour_submm} reveals that the alignment between reddening lanes and dust thermal 
emission is surprisingly poor. We have already identified some of the submm emission 
as hot-spots situated close to optically-bright HII regions. Thus some of the brightest regions 
at $850\mu$m may well correspond to warm grains heated by young stars (T=35K; Bianchi et al 2000b). 
In contrast, optical extinction is {\em not} sensitive, in any direct sense, to grain temperature 
although dust clouds situated in front of bright, blue star-forming regions will be detected quite 
readily in B-K maps. The situation contrasts strongly with the conclusions drawn for the nearby 
spiral NGC 891 (Alton et al 2000). Here, radial profiles along the major axis of this galaxy 
revealed a fairly good correspondance between optical reddening and $850\mu$m emission. Furthermore, 
there were strong indications from the thermal spectrum of NGC 891 that the submm is dominated by 
cold grains (15-20K) rather than dust surrounding HII regions (Alton et al 1998b). 
One possible explanation is that diffuse emission is dominating in NGC 6946, and remains undetected
because of the face-on orientation (while it is detected in the edge-on NGC 891).
Radial profiles of NGC 6946, to be discussed in \S\ref{emissivity}, 
suggest a stronger link between visual optical depth and $850\mu$m surface brightness than might be 
surmised from fig.~\ref{colour_submm}.

In addition to accentuating the absorption lanes, our B-K colour map highlights the 
presence of strong reddening towards the centre of NGC 6946. This is accompanied by an 
increase in $850\mu$m surface brightness; both properties, taken together, suggest an 
increase in dust concentration towards the galaxy nucleus. For the analysis that follows, 
we assume that the reddening in NGC 6946 is solely attributable to extinction by dust 
(and {\em not}, for example, caused by a colour change in the underlying stellar population). 
We adopt the observed B-K=3.1 colour at a radius of $4.7\arcmin\,$ ($R\simeq 0.9 \times R_{25}$) as the 
intrinsic, unreddened colour.
 This radius of $4.7\arcmin\,$ is chosen because extinction is small there, and it is the
largest radius where colours can be estimated properly.
We then assume that any increase in the colour index with radius ($\Delta(B-K)$)
arises from increasing optical depth.  The results in radial variation
(see fig. 8) do not {\it a posteriori} show the need for a more sophisticated
approach.
A Monte Carlo radiative transfer model is then employed to 
infer the visual optical depth, $\tau_{V}$, from $\Delta(B-K)$.\\

The assumption that we are measuring the true, underlying colour at a radius of $R=0.9 \times 
R_{25}$ is fairly robust -- the optical depth is generally believed to be quite low near the 
$R_{25}$ ($\tau_{V}\simeq 0.2$; Alton et al 2001). However, our assumption that an increase in 
$\Delta(B-K)$ is purely attributable to dust is probably incorrect, although this viewpoint 
has been proposed by many previous authors (Peletier et al 1995; Beckman et al 1996). 
Peletier et al (1995) find a statistical increase in the ratio of B-band radial scale-length to 
K-band radial scale-length for spiral disks more inclined to the line-of-sight. This effect is 
most readily explained in terms of dust reddening because the ratio would not change if the 
scale-lengths were determined by the stellar population gradient. In contrast, for a sample of 
86 face-on spirals, de Jong (1996) concluded that the change in scale-length across several 
optical and NIR passbands could not be understood without gradients in star-formation and 
metallicity being considered as a major factor. The assumption in this paper that increased B-K 
colour originates purely from increased optical depth will lead us to overestimate the opacity 
of the NGC 6946 disk. Furthermore, since the stellar population along the spiral arms is usually 
considered younger, and therefore bluer, than adjacent interarm regions, we will tend to both 
overestimate the amount of interarm dust and, therefore, overlook the propensity for dust to 
clump along the spiral arms (Knapen \& Beckman 1996).\\

\begin{figure}[t]
\centering
\resizebox{\hsize}{!}{\includegraphics{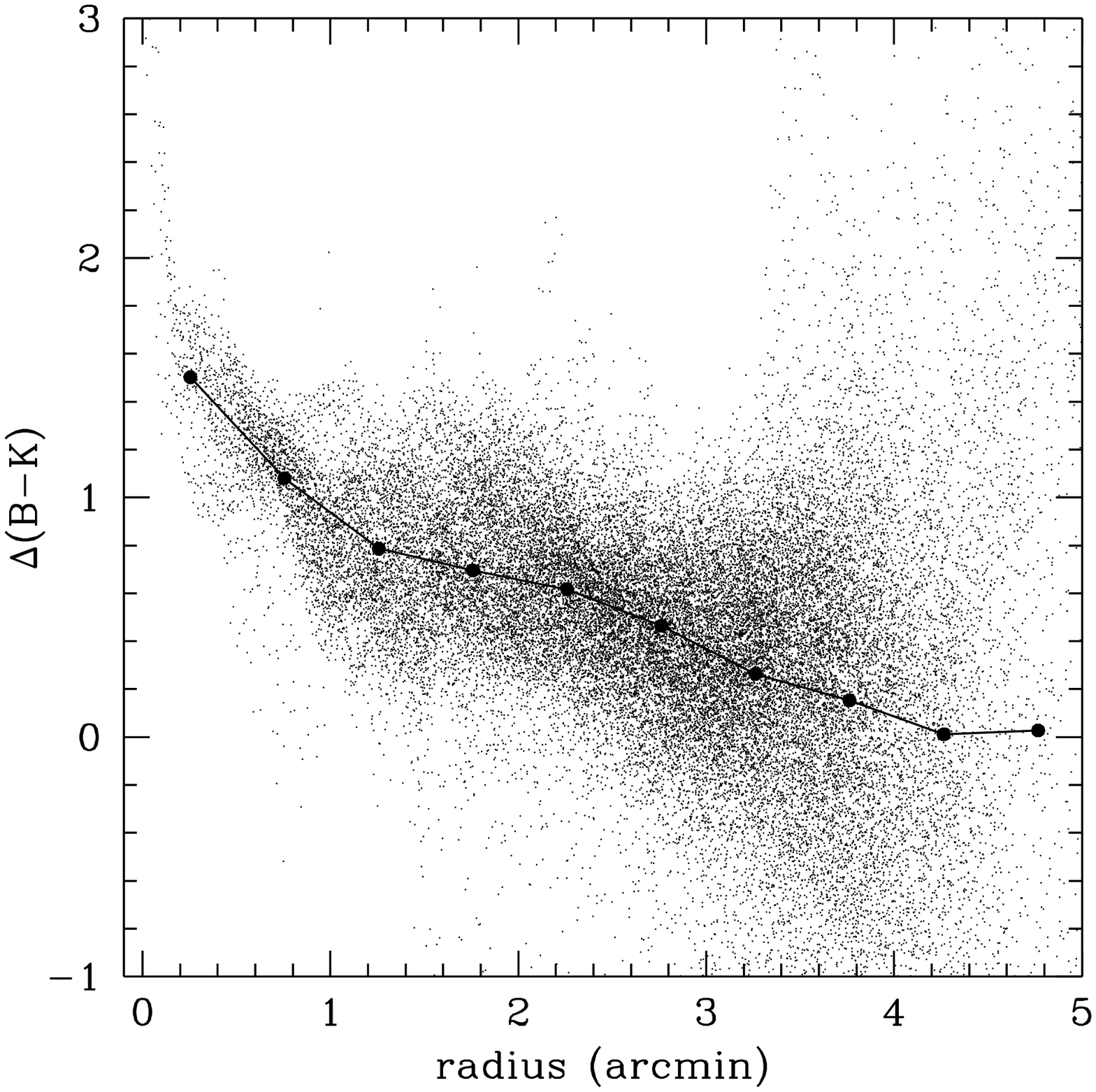}}
\caption{A breakdown of the B-K colour in fig.~\ref{colour_submm}. Dots represent the difference in 
colour of each pixel, $\Delta(B-K)$ with respect to the colour at the disk edge  ($R=4.7\arcmin\,$). The 
baseline at $R=4.7\arcmin\,$ has been established by means of an azimuthally-averaged profile (solid circles
 and solid line).}
\label{colour_prof}
\end{figure}

Fig.~\ref{colour_prof} provides a breakdown of the reddening measured in fig.~\ref{colour_submm} 
on a pixel-by-pixel basis. The azimuthally-averaged radial profile in B-K colour has been superimposed.
 We interpret $\Delta(B-K)$ by means of the Monte Carlo radiative transfer (RT) model developed by
 Bianchi et al (1996). This simulation assumes smooth distributions of both stars and dust which 
decline exponentially with both increasing galactocentric radius and height above the midplane. The
 light paths of optical and NIR photons are followed, through a succession of scattering interactions,
 until absorption takes place or the photon in question exits from the galaxy. By comparing the number
 of photons that escape from each part of the object with the quantity emitted by stars in the 
underlying disk, it is possible to relate observed colour to visual optical depth. \\

For NGC 6946, we adopt a `standard geometry' with dust grains extending to half the scale-height 
of the stellar disk and the dust radial scale-length fixed at 1.5 times the stellar scale-length.  
Acceptable variations in 
the scale-height and scale-lengths of dust with respect to the stars (Trewhella 1997; Bianchi 1999) 
lead to uncertainties of $\simeq 33$\% in the implied visual optical depth over most of the disk 
($\sim 50$\% for the central arcminute). 
An $R^{1/4}$ bulge was added to the stellar light so that, in total, 
bulge stars contributed $\frac{1}{10}$ of the total optical+NIR luminosity of the galaxy. Finally, 
it should be noted that no account is taken of dust clumping in the model which normally would lead 
us to underestimate the true optical depth by 50-100\% (Witt \& Gordon 2000; Bianchi et al 2000c). \\

Rather than attempt a {\em global} fit between the observed and simulated B-K colours, each 
pixel in fig.~\ref{colour_submm} was treated independently. A series of simulations were run, 
with the face-on, visual optical depth through the centre of the galaxy model varying between 2 
and 10. Overall, the technique worked well; a notable exception being within $45\arcsec\,$ of the 
nucleus where the observed $\Delta(B-K)$ was too high to give a reliable indication of the underlying 
optical depth (B-K effectively saturates as $\tau_{V}$ approaches $\sim10$).\\

Using the method outlined above a map of visual optical depth was created to the same spatial 
resolution as our K-band image ($4.2\arcsec\,$ FWHM). To compare with 
the SCUBA image (next section), the $\tau_{V}$ map was then smoothed to $18\arcsec\,$,
yielding a central, face-on opacity of $\tau^{0}_{V}=7.6$ and an exponential scale-length of 
$98\arcsec\,$, close to that inferred from ISO $200\mu$m images. 
We conclude that the visual optical depths derived from our RT model are reasonable.\\

\begin{figure}[t]
\centering
\resizebox{\hsize}{!}{\includegraphics{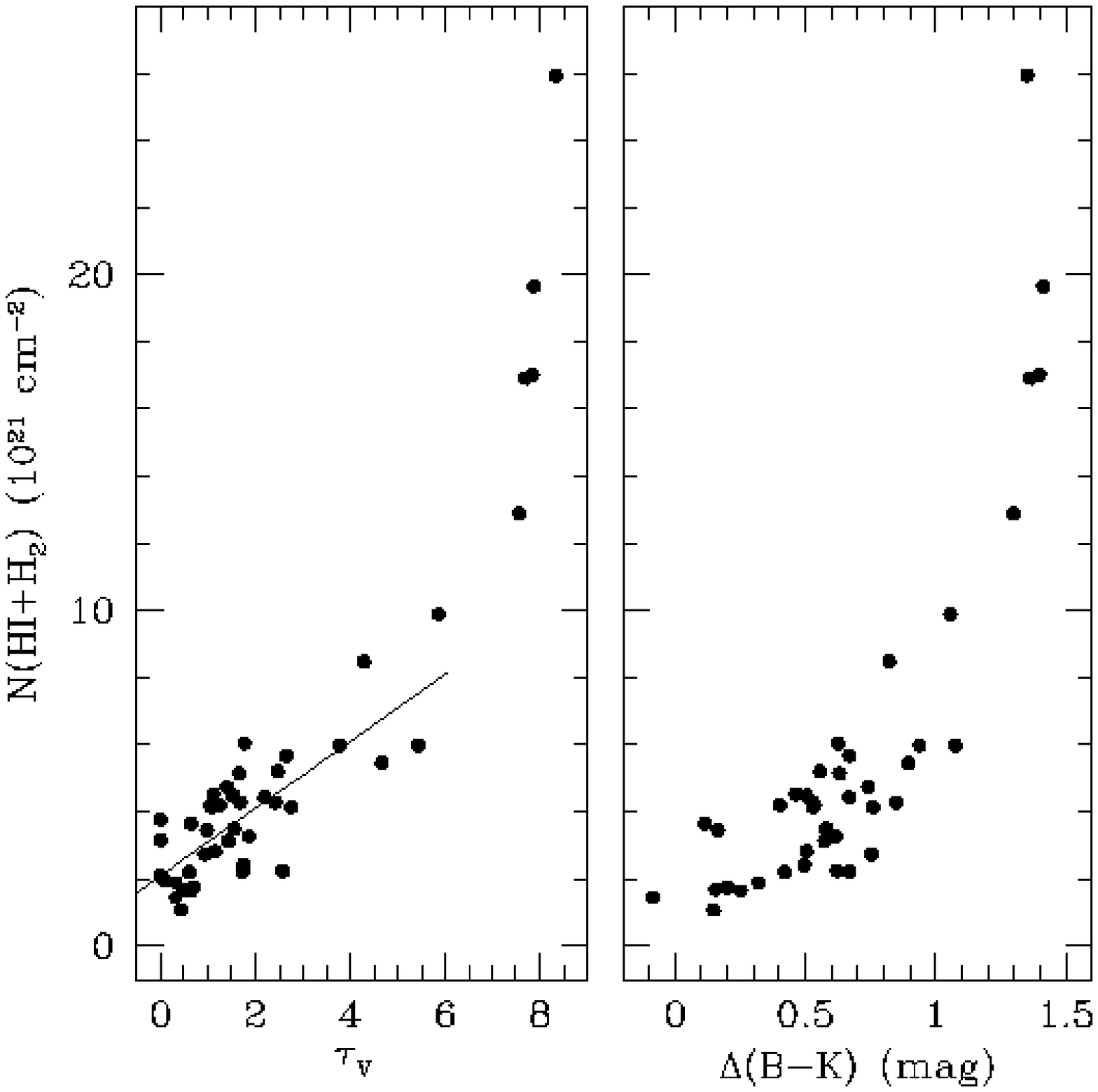}}
\caption{Neutral hydrogen column density (H$_{2}$+HI) plotted against both B-K colour and visual 
optical depth, $\tau_{V}$. The latter has been inferred from a Monte Carlo radiative transfer 
simulation. A least-squares linear fit between gas column density and $\tau_{V}$ is represented by
 the solid line (excluding $\tau_{V}>7$). The corresponding relationship in the Milky Way is also 
linear but with a gradient approximately twice that shown for NGC 6946.}
\label{gas_tau}
\end{figure}

Fig.~\ref{gas_tau} shows how the column density of neutral hydrogen in NGC 6946, $N(HI+H_{2})$, 
relates to both reddening, $\Delta(B-K)$, and the optical depth derived from RT modelling, $\tau_{V}$.
 If we exclude central regions saturated in B-K (and corresponding to  $\tau_{V}> 6$), 
a strong linear relationship emerges as follows
(correlation coefficient 0.91): 

\begin{equation}
\label{eq2}
\frac{N(HI+H_{2})}{10^{21}} = (1.0\pm 0.1) \tau_{V} + (2.1\pm 0.3)
\end{equation}

\noindent This is half the corresponding value measured within the solar 
neighborhood ($N(HI+H_{2})=2.1 \times 10^{21} \tau_{V}$; Bohlin et al 1978),
and could be due to the difference in metallicity. 
To obtain a total dust mass, we now convert $\tau_{V}$ into dust column density and integrate 
over the whole disk. To do this, we 
employ the relation given by Xilouris et al (1997) corresponding to `classical' Milky Way grains 
of size $0.1\mu$m and material density 3 g cm$^{-3}$ (Hildebrand 1983). Thus:

\begin{equation}
\label{eq3}
M_D = {\int} 0.27  \tau^{0}_{V}(R)  2  \pi  R  \hspace{0.1in} dR
\end{equation}

\noindent where the integral is over radius $R$, in metres, along the disk. Here, $\tau^{0}_{V}(R)$
 is the face-on optical depth at $R$ and $M_D$ is the total dust mass in g. Integrating 
eq.~\ref{eq3} out to a radius $R \sim 4\arcmin\,$, we derive $M_D=8.6 \times 10^{7} M_{\odot}$ 
for NGC 6946.
 Integrating $N(HI+H_{2})$ over the same region leads us to a global gas-to-dust ratio of 90. This
 value is consistent with the somewhat shallower gradient derived in eq.~\ref{eq2} compared with the 
solar neighborhood. Given our fundamental assumption in the derivation of $\tau_{V}$ (that B-K 
increases due to extinction only), our estimate of the gas-to-dust ratio must be considered a lower
 limit.

 Let us note that this value of 90 for the gas-to-dust ratio is about a factor 2 lower
than the average of 200 found towards 7 spiral galaxies (including NGC 6946) by Alton et al (1998a).
This is due essentially to the lower X=H$_2$/CO conversion ratio adopted here (X=1.5 instead of
2.8 10$^{20}$ cm$^{-2}$ (K km s$^{-1}$)$^{-1}$). Also the 850$\mu$m SCUBA emission could trace 
even colder dust than the ISO 200$\mu$m emission used by Alton et al (1998a).  For comparison,
the gas-to-dust ratio for the solar neighborhood is 150, and 100-300 for NGC 891,
according to the adopted conversion ratio or dust emissivity (Gu\'elin et al. 1993, 
Alton et al 2000).

\section{Inferred submm emissivity}
\label{inferred_emissivity}

The submm emission is not only a function of the dust column density and the dust 
temperature, but also of the nature of grains, which is reflected in their 
emissivity Q at various wavelengths.

A comparison between the SCUBA image and the optical depth just derived from RT modelling should
 allow us to constrain the $850\mu$m emissivity of dust grains in NGC 6946, 
$Q(850{\mu}m)$. The mass absorption coefficient ($\kappa$), which is often used, is related
 to the grain emissivity as follows: $\kappa_{850{\mu}m} = \frac{3Q(850{\mu}m)}{4a{\rho}}$, where
 $a$ and $\rho$ are the radius and material density of the grains respectively 
(Hildebrand 1983). It can be shown 
(e.g. Alton et al 2000) that optical depth and submm brightness are related as 
follows:

\begin{equation}
\label{eq4}
\tau^{0}_{V} = \frac{Q(V)}{Q(850{\mu}m)} \frac{1.67 \times 10^{-18}}{B(850{\mu}m,T)} \hspace{0.1in}
 cos(i) \hspace{0.1in} F_{850}
\end{equation}

\noindent where $\tau^{0}_{V}$ is the face-on optical depth, $Q(V)$ is the emissivity at 
$\lambda=0.55{\mu}m$, $i$ is the inclination of the disk ($i=90^{\circ}$ for edge-on), $F_{850}$ 
denotes the surface brightness of the $850\mu$m map, in Jy/$18\arcsec\,$beam, and $B(850{\mu}m,T)$ refers 
to the blackbody intensity at $850\mu$m for grains of temperature $T$. We can see that an estimate of 
$\frac{\tau_{V}}{F_{850}}$ would serve as a handle on $\frac{Q(V)}{Q(850{\mu}m)}$.\\

With current observational limits there exists almost an order of magnitude uncertainty 
in $Q(850{\mu}m)$ (and hence $\frac{Q(V)}{Q(850{\mu}m)}$, since $Q(V)$ is relatively well constrained
 to $\sim 1.5$; Whittet 1992). Indeed, emissivity values for wavelengths close to 1 mm are often 
extrapolated from observations of Galactic reflection nebulae near 100-200$\mu$m (see Alton et al 2000
 for a full discussion). At the same time we cannot overstate the importance of determining the 
submm/mm emissivity because, once $Q(\lambda)$ is known, submm/mm observations can then be furnished
 with visual optical depths in a relatively straightforward manner (Hildebrand 1983; Hughes et al 
1993).\\

\begin{figure}[t]
\centering
\resizebox{\hsize}{!}{\includegraphics{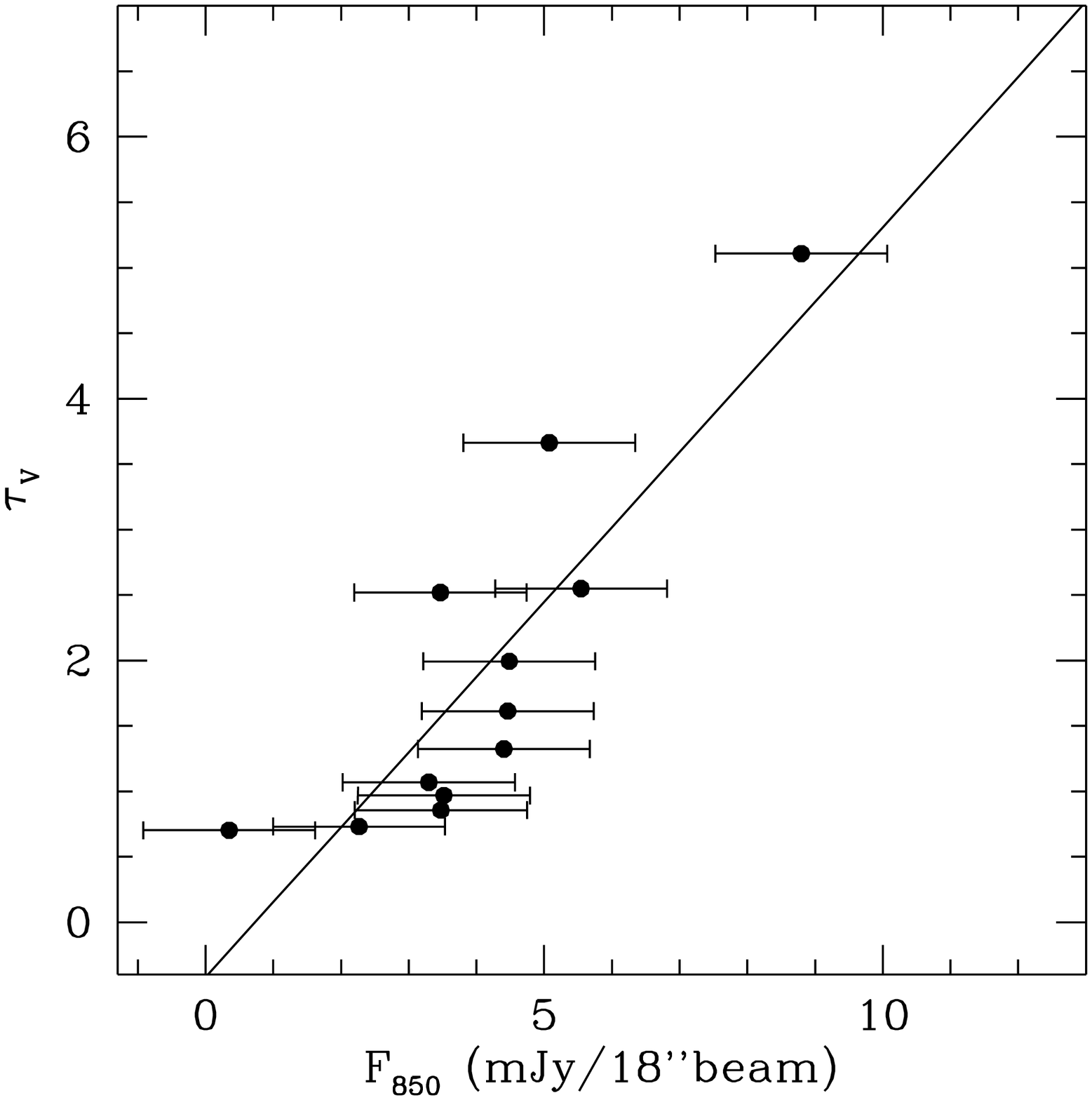}}
\caption{The relationship between visual optical depth (${\tau}_{V}$) and $850\mu$m surface brightness
 ($F_{850}$) for NGC 6946. Errorbars are only shown for $F_{850}$ as the values for ${\tau}_{V}$ are
 almost certainly upper limits. The solid line shows a least-squares linear fit to the data (see text).}
\label{tau_submm_radial}
\end{figure}

Radial profiles were fitted to both SCUBA and visual optical depth maps (both after smoothing to a
 spatial resolution of $18\arcsec\,$). Values pertaining to the central arcminute of the object were
 discarded due to suspected saturation in the B-K colour. Fig.~\ref{tau_submm_radial} shows that
optical depth ${\tau}_{V}$ compares rather well to submm surface brightness $F_{850}$,
when azimuthally averaged (while it was not true locally, cf fig.~\ref{colour_submm}). 
A least-squares best fit between optical depth and submm brightness yielded the following:

\begin{equation}
\frac{{\tau}_{V}}{F_{850}} = 573 \pm 108
\label{eq5}
\end{equation}

 Given the already mentioned uncertainties on $\tau_V$ and $F_{850}$, this value of the ratio is
an upper limit.
If we are to derive $\frac{Q(V)}{Q(850{\mu}m)}$ from eqs.~\ref{eq4} and \ref{eq5} we must have 
some notion of the grain temperature $T$. Unfortunately, no signal was recorded from the main disk
 of NGC 6946 during our $450\mu$m SCUBA observations (which would have allowed us to plot the thermal
 spectrum). ISO $200\mu$m images of NGC 6946, in conjunction with IRAS $100\mu$m HiRes data, suggest
 dust temperatures close to 20K in the main disk. RT modelling by Bianchi et al (2000a) points to $T$
 falling from 23K at the centre of NGC 6946 to 17K at the $R_{25}$ radius (similar to values proposed
 for the diffuse Galactic ISM; Mathis et al 1983; Draine \& Lee 1984; Reach et al 1995; Masi et al 
1995). Observations at both 450 and $850\mu$m for the nearby spirals NGC 891 and NGC 7331 suggest that
 90\% of galactic dust maintains a temperature between 15 and 20K (Alton et al 1998b; Alton et al 2001;
 see also Gu\'{e}lin et al 1993). To translate $\frac{\tau_{V}}{F_{850}}$ into 
$\frac{Q(V)}{Q(850{\mu}m)}$ we adopt $T=18K$. We note that a 50\% uncertainty in $T$ will produce a
 50\% variation in $\frac{Q(V)}{Q(850{\mu}m)}$ since $F_{850} \propto T$ in the Rayleigh-Jeans tail
 of the spectrum. The substitution of temperature into eq.~\ref{eq4} yields $\frac{Q(V)}{Q(850{\mu}m)}
 \simeq 1.4\times 10^{5}$ and a mass absorption coefficient of $\kappa(850{\mu}m) \simeq 0.26$ 
cm$^{2}$g$^{-1}$ is we assume $Q(V)=1.5$ and grains of radius $0.1\mu$m and material density 
3  g cm$^{-3}$.\\

\begin{figure}[t]
\centering
\resizebox{\hsize}{!}{\includegraphics{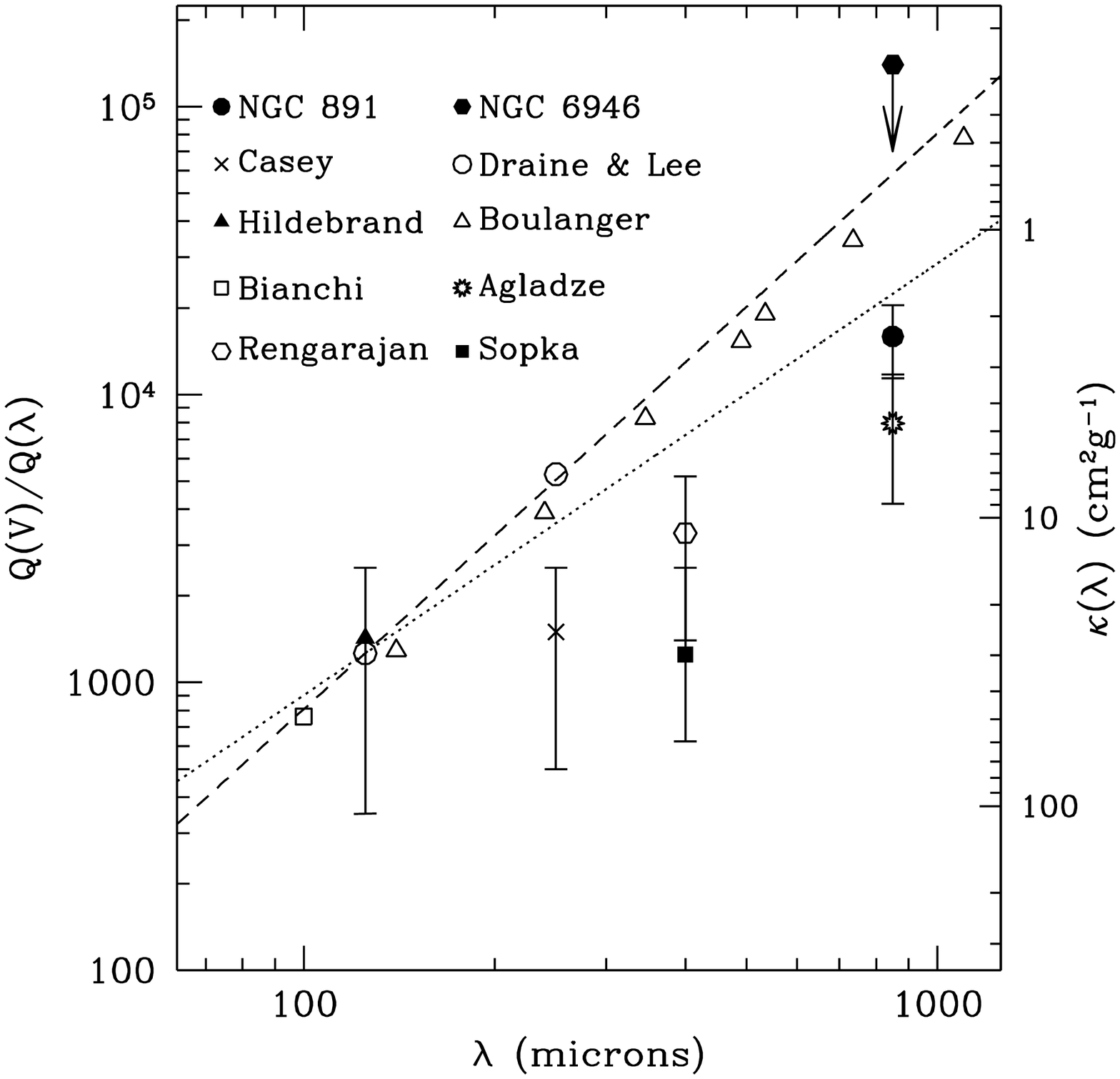}}
\caption{The emissivity of dust grains in the FIR/submm waveband. Measurements are plotted for 
Galactic reflection nebulae (Hildebrand 1983; Casey 1991), deeply-embedded FIR sources (Rengarajan
 1984), dust envelopes of evolved stars (Sopka et al 1985), the edge-on spiral NGC 891 (Alton et al
 2000) and NGC 6946 (upper limit, corresponding to our lower limit on submillimeter
emissivity given in this paper). Agladze et al (1994) refers to laboratory experiments
 conducted on silicate particulates. Draine \& Lee (1984) corresponds to a model for diffuse Galactic
 dust partly based on empirical data. We also show values derived from COBE and IRAS data based on
 emission from high-lattitude Galactic dust (Boulanger et al 1996 and Bianchi et al 1999 respectively).
 The dotted and dashed lines denote, respectively, $Q(\lambda) \propto \lambda^{-1.5}$ and $Q(\lambda)
 \propto \lambda^{-2.0}$ and are arbitrarily positioned to pass through the Draine \& Lee $125\mu$m
 point. The right hand axis shows the equivalent mass absorption coefficient ($\kappa$) assuming a
 radius and material density of $0.1\mu$m and $3$ gcm$^{-3}$, respectively, for the dust grains and
 $Q(V)$=1.5
}
\label{emissivity}
\end{figure}

The implied $850\mu$m emissivity is a factor 3 lower than that measured by COBE for diffuse 
Milky Way dust (suggesting we detect too little $850\mu$m radiation compared with the optical 
depth inferred from RT modelling). However, we recall that our $\tau_{V}$ values are only upper 
limits as we assumed from the outset that B-K only changes due to extinction. The aforementioned 
$850\mu$m emissivity should therefore be considered a {\em lower} limit. 
Also, there might be an underestimation of the emissivity, due to the possible 
existence of spatially flat diffuse emission
and the uncertainty in the background subtraction. In NGC 891, 
the submm emissivity was estimated $9\times$ higher than that for NGC 6946 (Alton et al 2000),
which might be due to more thorough detection of the diffuse
emission, because of edge-on orientation.

\begin{figure}[t]
\centering
\resizebox{\hsize}{!}{\includegraphics{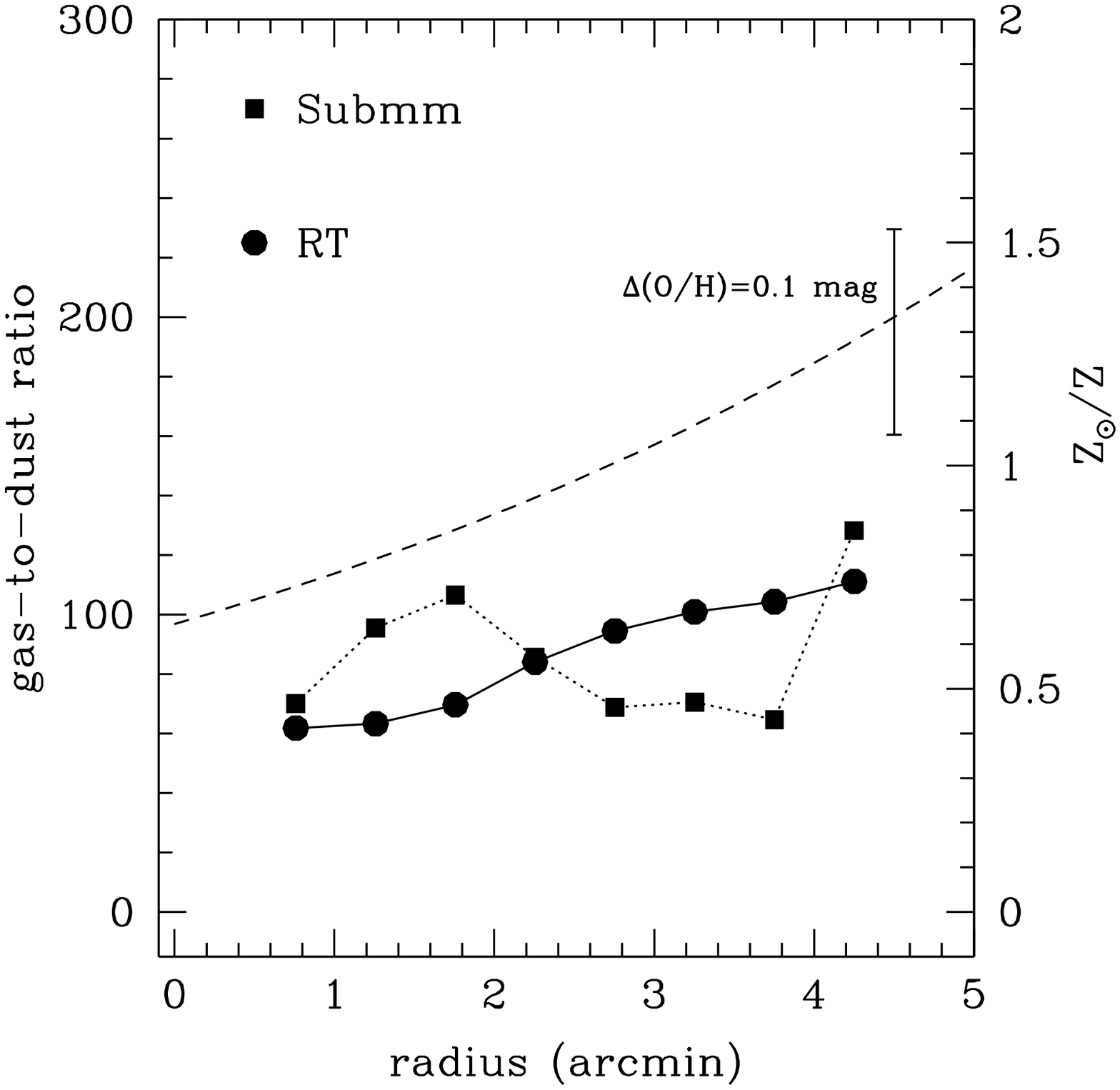}}
\caption{The radial behaviour in gas-to-dust ratio (left axis) with the dust mass derived from ;
 (i) RT modelling (solid line) and (ii) $850\mu$m emission (dotted line). The dashed line shows 
the gas-phase metallicity in NGC 6946 and its associated error (right axis, plotted as 
$\frac{Z_{\odot}}{Z}$). The vertical axes have been aligned so that a gas-to-dust ratio of 150
 coincides with solar metallicity. A vertical error-bar shows the probable systematic error in the 
metallicity measurement (Vila-Costas \& Edmunds 1992).}
\label{gd_metal}
\end{figure}

\section{Gas-to-dust Ratio}
\label{gd}

Fig.~\ref{gd_metal} reveals how the derived gas content varies with respect to
dust, with distance from the nucleus.  

In one case, the optical depth derived from RT modelling has been used to infer the dust 
mass per unit area (via eq.~\ref{eq3}). The gas surface density includes both atomic and molecular 
gas (the latter based on $X=1.5 \times 10^{20}$cm$^{-2}$ K kms$^{-1}$). The radial gas-to-dust ratio
 has also been calculated using the submm profile discussed in \S\ref{inferred_emissivity} and 
converting $F_{850}$ to $\tau_{V}$ with eq.~\ref{eq5}. The normalisation of $F_{850}$ to $\tau_{V}$
 explains why the mean gas-to-dust ratio indicated by both gas-to-dust profiles in fig.~\ref{gd_metal}
 is more or less the same.  Again, it must be emphasized here that
both curves are only lower limits.\\

We should remind here that the molecular gas
is dominating in NGC 6946, and its amount is estimated from the CO emission, which is itself
dependent on metallicity. In NGC 891 for example, both CO emission and submillimeter
dust emission have the same dependence with radius, suggesting that the CO
emission falls also proportionally to the metallicity (Gu\'elin et al. 1993).

A dashed line in fig.~\ref{gd_metal} represents the gas-phase metal abundance as measured 
from emission-line spectra (Vila-Costas \& Edmunds 1992). The vertical axes in the plot have 
been aligned so that solar metallicity corresponds to a gas-to-dust ratio of 150. The proximity 
of our gas-to-dust estimates to the dashed line is probably fortuitous given the factor 3-4 
uncertainty in deriving the former ($X$ factor, CO(2-1)/CO(1-0) etc.). 

Nearly no radial variation of the gas-to-dust ratio is observed. This means that the CO tracer
is also dependent on metallicity, in a similar way than the dust tracer.

\section{Dust and gas within arm and interarm regions.}
\label{arm_interarm}

The arm-interarm contrast in dust and gas is difficult to determine because of
limited spatial resolution
-- at $60\mu$m, an IRAS beam of $\sim 1\arcmin\,$ only begins to resolve spiral disks that are closer 
than $\sim 5$ Mpc. SCUBA, with a beam of $16\arcsec\,$ at $850\mu$m, can probe spiral arms separated 
by $1\arcmin\,$ (such as those in NGC 6946). However, the problem becomes one of sensitivity, 
with face-on disks possessing surface brightnesses of only a few mJy.\\ 

Although our 21cm HI observations are somewhat too coarse for the task ($25\arcsec\,$ FWHM), we have 
attempted a decomposition of the gas and dust associated with the spiral structure in NGC 6946, 
concentrating on the NE where the $850\mu$m emission is brightest. 
A series of 8 cross-sections across the arm-interarm interface were produced, sampling the opacity 
as well as the submm, CO and 21cm HI emission. The transects were then stacked, assuming an
average arm separation of $60\arcsec\,$ ($\lambda$) and 
stretching or compressing to take into account the variation of $\lambda$ with galactocentric radius.\\

\begin{figure}[t]
\centering
\resizebox{\hsize}{!}{\includegraphics{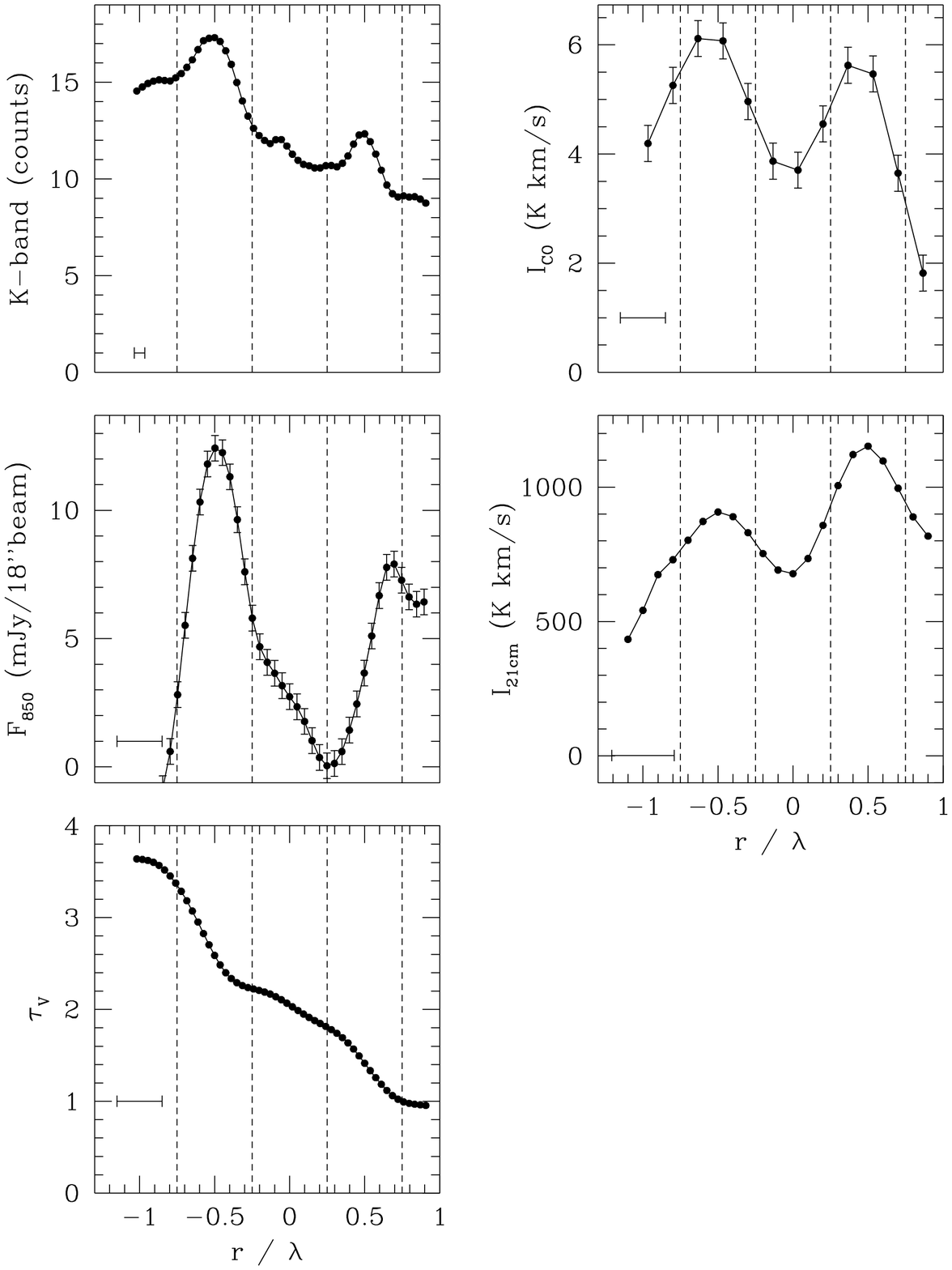}}
\caption{Profiles of gas and dust across the arm and interarm regions of NGC 6946. In all 
cases the horizontal axis denotes the distance from the interarm centroid ($r$) in terms of 
the mean arm-to-arm separation ($\lambda$). Dashed vertical lines delineate both the arm regions 
($r=-0.75\lambda$ to $r=-0.25\lambda$ and $r=+0.25\lambda$ to $r=+0.75\lambda$) and the interarm 
region ($r=-0.25\lambda$ to $r=+0.25\lambda$) as defined on the basis of the K-band image (top-left).
 Profiles are shown for $850\mu$m surface brightness ($F_{850}$), visual optical depth ($\tau_{V}$),
 $^{12}$CO(2-1) intensity ($I_{CO}$) and intensity in the 21cm HI line ($I_{21cm}$). Errorbars for these
 quantities are only shown when the uncertainties are much larger than the plotted markers. The spatial
 resolution of the data is represented by a horizontal error-bar at the bottom-left of each panel.}
\label{profiles}
\end{figure}

Fig.~\ref{profiles} depicts the mean profiles generated in this way. 
The vertical dashed lines define arm and interarm regions on the basis of the K-band profile 
(believed to trace the underlying stellar mass density). 
All the tracers of dust and gas plotted in 
fig.~\ref{profiles} show some modulation with the spiral density wave except perhaps the optical 
depth derived from RT modelling. 
 The lack of variation here probably stems from the assumption that an identical
 intrinsic B-K colour exists for the underlying stellar population in both arm and interarm regions.
 In reality, the arms will tend to be inherently bluer on average, due to the presence of young stars,
 leading us underestimate the arm opacity with respect to the interarm value.
The two forms of neutral hydrogen seem complementary in the sense that a reduced presence
 of HI along one arm is balanced by an enhanced level of H$_{2}$ and {\em vice versa}. \\

\begin{table}[h]
\caption{Properties of the arm and interarm regions in NGC 6946. Measurements are shown for the 
following: K-band light (row 1), $850\mu$m surface brightness (row 2), visual optical depth (row 3),
 $^{12}CO(2-1)$ intensity (row 4), H$_{2}$ column density (row 5), intensity in the 
21cm HI line (row 6),
 HI column density (row 7), total hydrogen column density (row 8), ratio of atomic to molecular 
column density (row 9), visual optical depth inferred from H$_{2}$ (row 10), visual optical depth 
inferred from HI (row 11), visual optical depth inferred from H$_{2}$+HI (row 12) and visual optical
 depth inferred from the $850\mu$m surface brightness (row 13). The H$_{2}$ column density assumes 
$X=1.5 \times 10^{20}$cm$^{-2}$ K kms$^{-1}$ and 0.4 for the ratio of CO(2-1) to CO(1-0) intensity.
 For inferences of visual optical depth see text. Due to the cancellation of systematic errors, 
uncertainties in the interarm-to-arm ratios are sometimes proportionally smaller than those pertaining
 to the arm or interarm region alone. Note that {\em none} of the quantities appearing in the table
 have been corrected to face-on values.}
\label{tab2}
\begin{tabular}{lllll}

Quantity 	& 	Units		&   Arm		& Interarm 	& Interarm  \\ 
		&			& average 	&  average 	&  / Arm \\ 
&&&& \\ 
K-band		&	counts		& $13.4\pm1.2$	& $11.4\pm1.2$	& $0.86\pm.02$ \\ 
$F_{850}$	&	mJy/$18\arcsec\,$	& $6.4\pm1.5$	& $2.7\pm1.5$	& $0.42\pm0.3$ \\ 
${\tau}_{V}$	&			& $\leq 2.05$	& $\leq 2.04$	& $\leq 1.0$	\\ 
$I_{CO}$	&	K km/s		& $5.32\pm0.4$	& $4.10\pm0.4$	& $0.77\pm.07$	\\ 
$N(H_{2})$	&$10^{21}$cm$^{-2}$     & $2.00\pm0.1$ & $1.54\pm0.1$ & as above \\ 
$I_{21cm}$	&     K km/s	 	& $941\pm58$	& $787\pm58$	& $0.84\pm.05$	\\ 
$N(HI)$		&$10^{21}$cm$^{-2}$	& $1.79\pm0.1$ & $1.50\pm0.1$ & as above \\ 
$N(H_2$+HI)&	$10^{21}$cm$^{-2}$	& $3.8\pm0.2$   & $3.0\pm0.2$   &  $0.78\pm.07$ \\ 
$\frac{N(H_{2})}{N(HI)}$	&	& $1.12\pm0.1$	& $1.03\pm0.1$	&	\\ 
${\tau}_{V}(H_2)$ 	&		& $0.96\pm.08$	& $0.74\pm.08$	&	\\ 
${\tau}_{V}(HI)$	&		& $0.86\pm.04$	& $0.72\pm.04$	&	\\ 
${\tau}_{V}(H_2$+HI)	&		& $1.8\pm0.1$	& $1.5\pm0.1$	&	\\ 
${\tau_{V}}(F_{850})$	&		& $3.7\pm0.9$	& $1.5\pm0.9$	&	\\ 

\end{tabular}
\end{table}

Table~\ref{tab2} contains estimates of the optical depth and gas column density based on the
 profiles presented in fig.~\ref{profiles}. The visual opacity is calculated variously from: 
radiative transfer (row 3); $850{\mu}$m surface brightness using eq.~\ref{eq5} (row 13); and the 
gas column density $N(HI+H_{2}$) (row12), the last of these assuming $\tau_{B}-\tau_{V} = 
\frac{N(HI+H_{2})}{6.3\times 10^{21} atoms/cm^{2}}$ as observed in the solar neighborhood (Bohlin 
et al 1978). Our estimate of $\tau_{V}$ based on RT modelling is expected to be an overestimate 
in the interarm region. Surprisingly, however, the gas column in such interstitial regions suggests
 equally high values of $\tau_{V}$. The submm observations favour quite a high interarm opacity as
 well ($1.5\pm 0.9$) but the combination of low signal-to-noise and difficulties in relating grain
 thermal emission to dust column density make the uncertainties here considerable. Attempts to
 determine interarm opacity by analysing the colour, surface brightness or number of background
 galaxies visible through a foreground spiral disk have generally led to values of less than unity
 for the face-on optical depth (White \& Keel 1992; White et al 1996; Berlind et al 1997; Gonzalez
 et al 1998).\\

Although the definition of arms and interarms regions is subjective,
we have attempted to compute the gas and dust mass in both regions.
We adopt 1.5 and 3.7, respectively, for the optical depth 
of the interarm and arm region at a radius of $2.3\arcmin\,$ (table~\ref{tab2}). In addition, we assume a 
radial scale-length of $98\arcsec\,$ for both arm and interstitial environments (\S\ref{reddening}). 
We suppose that interarm regions occupy only 50\% of the disk at $R=0.5 \times R_{25}$,
 increasing linearly with $R$ from 0\% at the nucleus to 100\% at the $R_{25}$ radius. Integrating 
eq.~\ref{eq3} suggests that 
at least 15\% of galactic dust is located between the spiral arms. A similar calculation for 
HI+H$_{2}$ (row 8, table~\ref{tab2}) indicates that at least 68\% of the neutral gas in NGC 6946 
can be considered as lying outside the spiral arms.

\section{Summary}
\label{summary}

We have analysed $850\mu$m SCUBA images of the nearby, (almost) face-on spiral galaxy NGC 6946. 
The submillimeter brightness level is low over most of the disk (often a few mJy at $850\mu$m) 
and, therefore, care has to be exercised in the subtraction of sky emission. Our main results can be 
summarised as follows:

\begin{enumerate}

\item Comparing the $850\mu$m flux with $^{12}$CO(2-1) and 21cm HI line intensity indicates a 
good correlation between dust thermal emission and {\em CO} emission. No correlation was 
detected between submm emission and {\em atomic} gas emission.

\item We place an upper limit on the visual optical depth in NGC 6946 ($\tau_{V}$) by means of a 
B-K colour map of the disk. We assume that the B-K colour at a radius $0.9 R_{25}$ corresponds to 
the underlying intrinsic colour of the stellar disk, in both arm and interarm regions, and that 
the observed increase in B-K with diminishing radius is attributable solely to extinction by dust. 
With these conditions we carry out a Monte Carlo radiative transfer simulation of the disk, 
inferring visual optical depth from B-K colour. 

\item The resultant map in $\tau_{V}$ relates fairly well to the distribution of neutral gas 
(HI+H$_{2}$). There is no significant radial variation of the derived gas-to-dust ratio,
which is easy to understand, given that the estimated gas content is dominated by the
molecular gas, therefore is estimated through the CO emission tracer. Since the CO emission is 
also dependent on metallicity, it is not possible to derive the true gas-to-dust ratio.
 This will await a large-scale mapping of other tracers, such as the pure rotational
lines of molecular hydrogen (e.g. Valentijn \& van der Werf, 1999).

\item Radial profiles of our optical depth map and SCUBA image indicate an association between 
$\tau_{V}$ and (diffuse) submm emission, $F_{850}$. A linear regression yields 
$\frac{\tau_{V}}{F_{850}} = 573 \pm 110$ and, by corollary, a $850\mu$m emissivity which is 
no more than 3 times lower than the COBE value for Galactic grains (Boulanger et al 1996). The
emissivity is also 9 times lower than in the edge-on galaxy NGC 891.
 This is however a lower limit, given the uncertainties on the visual
opacity, 
and the possibility of underestimating the large-scale submm emission. The
latter could be estimated in the future by more sensitive and 
more extended submm arrays.

\item A decomposition of spiral structure in NGC 6946 suggests an interarm optical depth of 
between 1 and 2 at a mean radius of $0.4 R_{25}$. These surprisingly high values represent 
40-80\% of the visual opacity along the arms. Furthermore, by a similar analysis, about two-thirds
of the neutral gas could be considered as lying outside the spiral arms.

\end{enumerate}

\begin{acknowledgements}
It is a pleasure to thank Gerald Moriarty-Schieven and Tim Jenness for their support during 
observations and data reduction. We are grateful to Steve Eales and Loretta Dunne for 
stimulating discussions on grain opacity and related matters, and to the referee for
his careful reading and positive criticism.
\end{acknowledgements}

\end{document}